# Situationally Induced Impairment in Navigation Support for Runners

**Shreepriya Shreepriya**
UX & Ethnography
Naver Labs Europe, Meylan, France
shreepriya.shreepriya@naverlabs.com

**Danilo Gallo**
UX & Ethnography
Naver Labs Europe, Meylan, France
danilo.gallo@naverlabs.com

**Sruthi Viswanathan**
UX & Ethnography
Naver Labs Europe, Meylan, France
sruthi.viswanathan@naverlabs.com

**Jutta Willamowski**
UX & Ethnography
Naver Labs Europe, Meylan, France
jutta.willamowski@naverlabs.com

## ABSTRACT
Mobile devices are ubiquitous and support us in a myriad of situations. In this paper, we study the support that mobile devices provide for navigation. It presents our findings on the Situational Induced Impairments and Disabilities (SIID) during running. We define the context of runners and the factors affecting the use of mobile devices for navigation during running. We discuss design implications and introduce early concepts to address the uncovered SIID issues. This work contributes to the growing body of research on SIID in using mobile devices.

**KEYWORDS**

Situationally induced impairments and disabilities; Mobility; Running; User Research; Navigation; Design framework

**INTRODUCTION AND RELATED WORK**

The impairment induced by the context of a user is defined as Situationally Induced Impairment [14]. The mobile state of the user constitutes an important factor for situational impairment with mobile devices by decreasing the usability [12]. Prior art has addressed this issue and proposed potential solutions by introducing interfaces that are more adapted to walking, e.g. by increasing the target size [7], introducing gestures [6] and using audio to give information to the user[19]. Similarly, wearable solutions are considered for specific situations, e.g. biking [8] and running [10].

The situation of mobility impairs the navigation support that is provided by mobile devices. Navigating on the move requires the users to divide their attention between their screen and the surroundings [19]. This deteriorates their experience of the environment and raises safety concerns. This is especially true for runners and cyclists, who, due to their increased speed compared to pedestrians, are in a more challenging situation with respect to navigation. They indeed have a limited time frame to comprehend the information provided by the mobile device and to make decisions based on that information. This situational impairment can be further deteriorated by several additional contextual factors, such as noise, light, and terrain conditions [12]. Moreover, the use of accessories in motion factor into the experience of situational visual impairment [18].

Traditional smartphone displays tend to be inefficient in this context. Indeed, users in motion do not wish to engage with their phone during their activity. They only bring them for their activity to stay reachable (by phone) or to get directions in case they get lost. Smartwatches are an alternate solution for navigation. They are easier to glance at than smartphones and some, e.g. Garmin [3] contain and display maps for navigation. However, the screen size is very small and thus the information displayed is limited. Accessing more detailed information requires additional movements and interactions like zooming in and out which are not suitable in motion.

There have been prior research efforts to bring freedom and safety in mobile navigation. Shoe me the way [13] used tactile feedback for providing turn-by-turn routing information to the user. NaviRadar [11] performed a radar sweep with a single vibrator attached to the back of a mobile phone. Commercial navigation support applications for smartphones and smartwatches guide the users through visual feedback. Products like Beeline Velo [1], Blubel [2] and SmartHalo [5] introduce new visual interfaces to redesign the experience of biking. VibroBelt [15] and BikeGesture [17] for instance use tactile cues and gesture inputs to help bikers to navigate. However, these works do not address the specific challenges posed by different contextual elements for runners and bikers.

The aim of our work is to solve the challenges of SIID in navigating with mobile devices during running. Prior art mainly focuses on the challenges of supporting pedestrian navigation. We focus on SIID during the activity of running. In this paper, we analyze the situational factors influencing the user while running, and discuss how these factors affect the usability of mobile devices for navigation.

## USER RESEARCH AND FINDINGS

When the user has to divide his attention among several tasks or elements while completing a main primary task, this has a negative effect on her/his cognitive performance [19]. In our case, we consider the user in the context of running with mobile navigation support as the primary task. To understand the elements that impact the user's running situation which require attention from the user, we conducted interviews with seven regular runners (5 men and 2 women, age between 30-55). This helped us to better understand all the contextual elements that impact the runners' behavior and the interactions runners (want to) have with their mobile devices during running. **Figure 1** illustrates runners and their context in terms of internal, i.e. user dependent factors, and external, i.e. user independent factors.

### Internal factors

We identified four factors that condition the navigation support needs of our participants during their running situation, (1) the *space* the user is running in, (2) the *company* the user runs with, (3) the *time* available, and (4) the expected and/or preferred *usage* of mobile devices during the run.

With respect to the *space*, the user may run anywhere *between completely familiar* and *completely unfamiliar* environments. To navigate in familiar spaces, the user simply looks around and searches for familiar cues and has no or little need for navigation support. To navigate in completely unfamiliar spaces, in contrast, very detailed and precise navigation support is required. Without appropriate navigation support many users avoid running in unfamiliar places altogether. In any case, the challenge is to provide the *desired* navigation support with minimal interruption to the running experience.

With respect to *company*, this factor corresponds to the fact that a user is running alone or together with other runners. For the purpose of this paper, we do not consider this aspect. When running in groups, people either navigate collaboratively, or they often rely on one person playing the role of the navigational guide for the group. In both cases, navigation support is less required and/or the situation is relatively similar to running in familiar environments.

With respect to time constraints, runners often want to practice their activity for a pre-determined distance or time. This impacts and interferes with the navigation task, as the user wants to get back to the pre-defined point in time and needs support to navigate there accordingly.

With respect to the usage of the mobile device during the running situation, in general runners try to minimize this usage and their interaction with the mobile device as this essentially requires interrupting the activity.

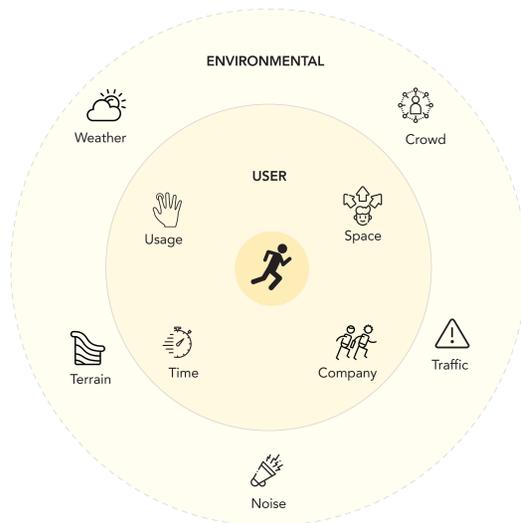

**Figure 1: Illustration of runner and context**

**External factors**

*External factors include (1) weather conditions, (2) crowds, (3) traffic, (4) noise and (5) terrain.*

*Weather* conditions like rain and strong sunlight can impede the ability of the runner to see the mobile screens. For example, in case of strong sunlight, runners generally place a covering hand between the mobile screen and their eyes to clearly see the displayed information. This increases the time required to stop and interact, thus breaking the flow of their running activity.

*Crowds*, *traffic*, *noise* and *terrain* also impact the running experience and the amount of spare attention available for navigation. Depending on their characteristics these elements demand the runner's attention. In the presence of crowds, the runner has to be more cautious to avoid the risk of bumping. Similarly, heavy traffic requires higher attention. The same applies for difficult terrain, where the user has to pay attention not to fall and hurt. Finally, in the presence of noise, it may be difficult to concentrate on the navigational cues provided by a mobile device. Thus, in these contexts, the user can only dedicate a limited amount of attention to the navigation task.

## IMPLICATIONS FOR DESIGN

Using mobile devices for navigation during a run constitutes an issue for the runner. It demands the user's attention that is taken away from the contextual elements. The understanding of these factors gives a perspective on the desired solutions that can help make users free to connect with the environment and people. A good design should be molded to demand minimal user attention. Below, we list some implications for design and conceptualize solutions for situational impairments while running, resulting from our analysis.

**Natural Interaction**
Runners need to stop or slow down and interact with the mobile device to get navigational cues. This interaction is often unnatural and requires additional movements like looking at a mobile phone. While glancing at smartwatches for navigational cues is less disruptive than looking at the screen of a mobile phone, it still constitutes an unnatural movement for a runner. Navigational solutions that make use of the runner's body movements and are more focused on natural interactions would be more appropriate and reduce the hazard of situational impairments.

**Un-intrusive**
Current audio and visual navigational cues are intrusive as they require the runner's attention to be focused on the screen or on voice commands. With their attention divided between the screen (or audio) and the environment, the runner is distracted from the environment and the provision of navigational information becomes significantly more difficult and less effective. Our aim is to provide navigation support in a more un-intrusive way. Also, information displayed on a screen is impacted by weather conditions and audio by ambient noise. We attempt to propose solutions that are *not* subject to such environmental conditions.

|  | Natural Interaction | Un-intrusive | Personalization |
|---|---|---|---|
| Concept 1: Each path has a sound | ●●●○○ | ●●●○○ | ●●●●○ |
| Concept 2: Follow the white tiger | ●●●●○ | ●●●○○ | ●●●●○ |
| Concept 3: Thumbs Up | ●●●●● | ●●●●○ | ●●●○○ |
| Concept 4: Let music be your guide | ●●●●○ | ●●●●○ | ●●●●● |

**Figure 2: Analysis of concepts based on design implications**

**Informed by Individual Preferences**
Each runner may have different preferences and needs with respect to navigation support. The optimal frequency of interaction with the mobile devices during the running activity is dependent on the individual's needs and preferences and on the context, e.g. how much knowledge the user has about the environment and whether she/he has planned out his tour path in advance or not. Solutions that can adapt to the individual, to her/his preferences and to the actual context can better contribute to reduce situational impairments.

**CONCEPTS**
In the following, we introduce a selection of concepts we developed during brainstorming sessions. We follow our design guidelines and analyze the effectiveness of each concept (**Figure 2**).
We explored four concepts, that take different approaches towards the problem of providing mobile navigation support for runners in unfamiliar places. They use technologies like augmented reality, sound or haptic feedback to effectively guide the users along a running path. What they have in common is that they all present adaptive, natural and non-intrusive user interfaces that avoid any additional or unnatural movements.

**Concept 1: Each Path has a Sound**
Runners are guided through sound, conveyed over earphones or similar audio devices. When reaching intersections and decision-making points they hear sounds coming from the different available pathways conveying the paths' characteristics (**Figure 3**). For example, birds chirping would indicate a nearby park in that direction whereas honking sounds of vehicles could denote busy roads. As previously studied, audio is a less intrusive way to communicate information when on the move [19], and the use of spatial sound can constitute an effective and intuitive guiding cue [16][9]. One of the challenges is to define the right sound effects to convey the information, and the right time to provide them to avoid disturbing the rhythm of the runner. These elements could be modified by the runners to make sure they are not annoying and adapted to the runner's needs.

**Concept 2: Follow the White Tiger**
A virtual character that guides and informs the runner about different route options.

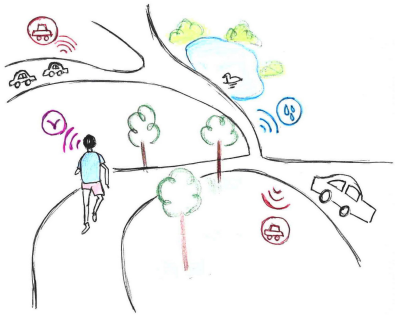
**Figure 3: Each Path has a Sound**

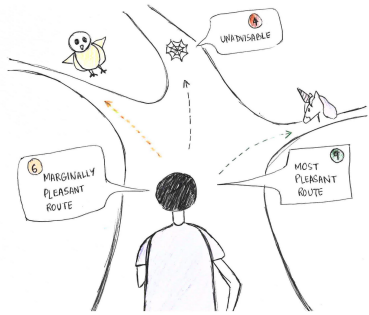
**Figure 4: Follow the White Tiger**

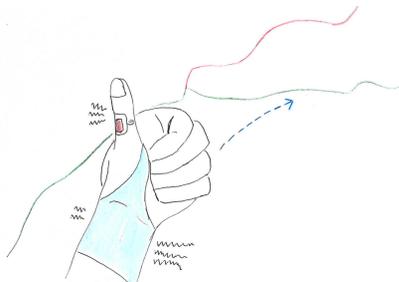
**Figure 5: Thumbs Up**

This idea is based on the insight that people like to follow or chase likeable characters [4]. The virtual characters are displayed to the runners through Augmented Reality glasses. Each path can be represented by a different character with characteristics that represent its quality (**Figure 4**). For example, a tiger can denote a challenging path. The characters as well as the frequency with which runners see them during the run can be defined and edited by each runner. They can choose to have a constant companion or one that shows up at specific points to guide them through the path. Even if using visual elements, this proposal constitutes an un-intrusive solution where the presented information merges with the environment. It does not compete for the attention required to navigate the surroundings or demand any additional movements from the user. The challenges associated with the concept are mostly technical: the AR glasses must be light enough to be comfortably carried by the runner and the positioning system precise enough to place the characters in the right spots.

**Concept 3: Thumbs Up**
The side view of the thumbs is the most natural thing to view for runners and their position is suitable to give quick guidance about upcoming paths at intersections. This can be capitalized by wearing a ring on each thumb and individually adapt their feedback, using either light or vibration, to direct the runners towards the right path (**Figure 5**). The characteristics of the vibration, as well as the distance from the intersection it is triggered can be defined by the runner. This device is an intuitive solution that doesn't require unnatural movements or much attention from the user. Even though this would be minimally intrusive, it could also turn into a weak point, since the cues could be missed or misunderstood [20], especially while performing an activity like running. User tests are needed to fully understand its feasibility as a navigation method.

**Concept 4: Let the Music be Your Guide**
According to our research, a majority of runners listen to music while exercising. This concept takes advantage of music, combining its modulation with natural head movements to guide runners towards the right paths. The concept is strong on all the design implications and was the one we selected to pursue with development. We produced an early prototype that has been tested with users and received positive feedback. We are currently working on the second prototype that improves on identified shortcomings of the previous version. However, in this paper we don't elaborate more on this due to confidentiality issues.

## CONCLUSIONS

In this study, we investigate the effect of mobility on mobile interactions in navigation. Through user interviews, we categorize affecting factors based on the context of the user and the external environment. The analysis of these factors led to our three-dimensional design guidelines of natural and un-intrusive interactions that are adaptive to user's preference to solve SIIDs. These guided our brainstorming session to propose potential solutions. Through our system we identified the most suitable solution that we are currently working on. Our findings enhance the understanding of situational impairments on the move focused on the task of navigation. They outline the research process and define a design framework that can be adopted for navigation tasks in other mobility scenarios.